\documentclass[12pt,a4]{article}                                               
\title{Group Theoretical Structure and Inverse Scattering
Method for super-KdV equation}
\author{ Petr P. Kulish$^1$, Anton M. Zeitlin$^2$\\
$^1$ St. Petersburg Department \\
of Steklov Mathematical 
Institute, Fontanka, 27,\\ 
St. Petersburg, 191023, Russia\\
$^2$ Department of High Energy Physics,\\ Physics Faculty, St. Petersburg State 
  University,\\ Ul'yanovskaja 1, Petrodvoretz, St.Petersburg, 198904, Russia\\}
\date{}

\begin{document}
\maketitle
\begin{abstract}
Using the group-theoretical approach to the inverse scattering me\-thod the
supersymmetric Korteweg-de Vries equation is obtained by appli\-cation
of the Drinfeld-Sokolov reduction to $osp(1|2)$ loop superal\-gebra.
The direct and inverse scattering problems are discussed for the
corresponding Lax pair.
\end{abstract}
\section*{Introduction}
This article is devoted to the study of supersymmetric extension of the 
Korteweg-de Vries equation
(super-KdV) ([4], [5]):
\begin{eqnarray}
u_t&=&-u_{xxx}+6uu_x-12\varepsilon\varepsilon_{xx},\\
\varepsilon_t&=&-4\varepsilon_{xxx}+6u\varepsilon_x +3u_x\varepsilon,\nonumber
\end{eqnarray}
where $\varepsilon(x,t)$ is a function with values in
the odd part of the Grassmann algebra  $\Lambda(n)_{\bar 1}$, 
and $u(x,t)$ is a function with values in its even part $\Lambda(n)_{\bar 0}$ 
(see Appendix). 
This equation is Hamiltonian one with the following Poisson brackets:
\begin{eqnarray}
\{u(x),u(y)\}&=&
\frac{1}{2}\Big(\delta'''(x-y)-2u'(x)\delta(x-y)-4u(x)\delta'(x-y)\Big)
\nonumber\\
\{u(x),\varepsilon(y)\}&=&
-\frac{1}{2}\Big(3\varepsilon(x)\delta'(x-y)+\varepsilon'(x)\delta(x-y)\Big)
\nonumber\\
\{\varepsilon(x),\varepsilon(y)\}&=&
-\frac{1}{2}\Big(\delta''(x-y)-u(x)\delta(x-y)\Big)\nonumber
\end{eqnarray}
which generate the superconformal algebra.
The corresponding Hamiltonian is:
\begin{eqnarray*}
H&=&\int(u^2(x)-4\varepsilon(x)\varepsilon'(x))dx.\\
\end{eqnarray*}
This equation is integrable and allows the Lax representation:
$\dot{{\cal L}}=[{\cal M},{\cal L}]$. 

The fist part of this article is devoted to the mentoned above, 
i.e. obtaining  the Poisson structure and the Lax representation for the
super-KdV equation.

In the second part the application of the inverse scattering method 
(ISM) \cite{6}-\cite{8} to this system is considered. 
The attention is devoted mostly to the case of 
one-dimensional Grassmann algebra,
when the super-KdV equation is reduced to the system of equations:
\begin{eqnarray}\label{eq:redsys}
u_t&=&-u_{xxx}+6uu_x\\
e_t&=&-4e_{xxx}+6ue_x\nonumber + 3u_x e
\end{eqnarray}
where $u(x,t), \ e(x,t)$ are some functions with values in ${\bf R}$.

This system naturally appears in the case of Grassmann algebra 
of any dimension,
if we write down the super-KdV equation using the basis of monomials:
$\Theta^{i_1} \Theta^{i_2}\dots\Theta^{i_n}$
 (see Appendix). Really, restricting to the terms, linear in $\Theta^i$ , 
we obtain the system (\ref{eq:redsys}).

Morover, it should be noted, that the only nonlinear equation  
(in the basis decomposition of the super-KdV equation) is the usual 
KdV equation, all others are linear and the whole system can be 
written in the ``triangular form''.\\ 
One-dimensional case is interesting, because the pair of equations 
(\ref{eq:redsys}) can be expressed in such a way:

\begin{eqnarray}
\left\{ \begin{array}{l}
\dot{L}=[L,A]\\
e_t=-Ae
\end{array}
\right.
\end{eqnarray}
where $L=-\partial^2_x+u(x), A=4\partial^3_x-3(u\partial_x+\partial_xu)$
is a scalar L-A pair for usual KdV.
One can easily derive that if $e(x,t)$ is a solution for given $u(x,t)$,
then $L^ne, n=1,2,\dots$ is also a solution for this $u$.
Thus we obtain an infinite chain of solutions, except the
case when $e(x,t)$ is an eigenfunction of $L$.
For example, if one takes the one-soliton solution of KdV: 
\begin{equation}
u(x,t)=\frac{c}{2{\rm ch}^2\Big(\frac{\sqrt{c}}{2}(x-ct)\Big)}
\end{equation}
and then supposing, that $e(x,t)$ is an eigenfunction of $L$,
one can obtain a solution:

\begin{equation}
e(x,t)=\frac{a}{2{\rm ch}^2\Big(\frac{\sqrt{c}}{2}(x-ct)\Big)}
\qquad (a \in {\bf R}).
\end{equation}
These solutions can be obtained also by the ISM (see part 2).
Nowadays many papers are devoted to the superequations due
to the progress in the superstring theory.
It is reasonable to mention the papers [9]-[11].
Moreover, if we want to quantize 
such systems, it would be useful
to study their behavior on the classical level.

\section{Group-theoretical structure\\ 
of the super-KdV equation}
The usual KdV equation 
is related with the scalar Lax operator
$$
L=-\partial_x^2 + \lambda + u
$$
or with the equivalent matrix Lax operator
$$
\partial_x+\left(\begin{array}{cc}
0 & 1\\
u+\lambda & 0
\end{array}
\right),
$$
and the Lie algebra $sl(2)$. 
The Hamiltonian structure for the KdV equation is given 
by the Virasoro algebra:
\begin{equation}
\{u(x),u(y)\}=\frac{1}{2}\Big(\delta'''(x-y)-(4u(x)\partial_x+
2u_x(x))\delta(x-y)\Big)
\end{equation}
The relation of these two objects with the KdV equation is given by
the group-theoretical scheme and the Drinfeld-Sokolov reduction.

In this part of the article, this formalism 
is applied to the superalgebra ${\rm osp}(1|2)$,
and following this approach the super-KdV equation is obtained.

\subsection{Group-theoretical structure and the \\
Drinfeld-Sokolov reduction for
${\rm osp}(1|2)$}
The following constructions were applied to the 
Lie algebras (see, for example 
[1],[2],[12]).
We extend them to the case of superalgebras, in particular to the
${\rm osp}(1|2)$ superalgebra (see Appendix).

Let ${\cal G}$ be a Lie superalgebra and let 
$$
A=\ell({\cal G})
={\cal G}\otimes C[\lambda,\lambda^{-1}]=\{\sum^s_{i=r}X_i\lambda^i|X_i\in 
{\cal G}\}
$$
be its affinization by the spectral parameter $\lambda$. $A=A_++A_-$, where
$A_+ ={\cal G}\otimes C[\lambda]=
\{\sum^s_{i=0}X_i\lambda^i|X_i\in 
{\cal G}\}$, 
$A_- ={\cal G}\otimes\lambda^{-1}C[\lambda^{-1}]= 
\{\sum^{-1}_{i=r}X_i\lambda^i|X_i\in 
{\cal G}\}$.
Let $R 
=P_+-P_-$, where $P_+$ and $P_-$ are projectors on $A_+$ и $A_-$ 
correspondingly.
Moreover, let us introduce multiplication operator  $\hat{\eta}:A\to A$
for function $\eta(\lambda) \in C[\lambda,\lambda^{-1}]$ in such a way: 
$(\hat{\eta}X)(\lambda)=\eta(\lambda)X(\lambda)$. Then we define
the operator
$R_\eta=R\circ\hat{\eta}$.
It appears that $R_\eta$ defines classical $R$-matrix for all 
$\eta$, i.e. 
$$
[X,Y]_{R_\eta}=\frac{1}{2}[R_\eta X,Y]+\frac{1}{2}[X,R_\eta Y]
$$
is also a Lie superbracket.

We can identify the dual space $A^*=\ell({\cal G})^*$ with 
$\ell({\cal G}^*)={\cal G}^*\otimes C[\lambda,\lambda^{-1}]$ 
by means of pairing:
\begin{eqnarray}
&&<\alpha,x>=<\sum_i\alpha_i\lambda^i,\sum_jx^j\lambda^j>=
\sum_{i+j=-1}\alpha_i(x_j), \\
&&\quad \alpha_i\in {\cal G}^*, \ x_i\in {\cal G}\nonumber
\end{eqnarray}
Now we are ready to introduce the Poisson brackets on  $A^*$ (we mean
structures of the Lie Poisson type).   
We are interested in the following Poisson brackets:
\begin{equation}
\{\phi,\psi\}_{R_\eta}(\alpha)=<\alpha,[\nabla_\alpha\phi,\nabla_\alpha\psi]_{R_\eta}>
\end{equation}
where $\phi, \psi \in C^\infty(A^*), \ \alpha\in A^*$.
If we consider the space of functions, invariant under the
coadjoint action: 
$$
I(A^*)=\{\phi\in C^\infty(A^*)|\phi(Ad^*_{g}\alpha)=\phi(\alpha), \qquad 
\forall g\in \ell(G) \},
$$
where $\ell(G)$ is a supergroup, constructed from the Lie superalgebra 
$\ell({\cal G})$, then the same proposition as for usual
Lie algebras takes place, i.e. 
$$
\{\phi,\psi\}_R(\alpha)=0, \qquad
 \alpha\in A^*,  \phi, \psi \in I(A^*)
$$
Now let's consider the central extension of our algebra. We will call 
 ${\cal G}_c$ 
the linear space of pairs
$\quad{\cal G}_c =\{(X,a)|X:S^1\to {\cal G};a\in {\bf C} \}$, with the 
following 
bracket:
$$[(X,a),(Y,b)]=(XY-YX,str\int_{S^1} dx  X'(x)Y(x)).
$$ 
One can obtain the isomorphism between
${\cal G}_c^*$ 
and the space of operators ${\cal L}=e\partial_x + \mu$, 
where $(\mu,e)\in {\cal G}_c^*$. The pairing here is defined as follows: 
$<(\mu,e),(X,a)>=ea+str\int\mu X$.
Using the previous discussion, we can introduce the affinization of 
${\cal G}_c$ by means of spectral parameter $\lambda$
and obtain in such a way algebra $A^*=\ell({\cal G}_c^*)$.
From the identification of
$\ell({\cal G}_c^*)$ with the differential operators one can easily see that
the coadjoint representation is given by the ``nonabelian gauge 
transformation'' [1]:
$$
{\cal L}=e\partial_x + \mu \to g{\cal L}g^{-1}=e\partial_x+ g\mu g^{-1}-eg'g^{-1},
$$
where $g\in \ell(G)$. 
Consider the space $H_{C}({\cal G})$, consisting of the
elements of the following type:
$(\mu,e)(\lambda)=(J(x)+C\lambda,e_0+e_1\lambda)$.
Let $\eta(\lambda)=\eta_0+\eta_1\lambda$.
Then $R_\eta$ Poisson bracket 
can be expressed in such a way:
$$
\{\phi,\psi\}_{R_\eta}(J)=
-\eta_0 \Big(str\int C \Big[\frac{\delta \phi}{\delta J(z)},
\frac{\delta \psi}{\delta J(z)}\Big]
$$
$$+
e_1 str\int\partial_z\Big(\frac{\delta \phi}{\delta J(z)}\Big)
\frac{\delta \psi}{\delta J(z)}\Big)+
$$
$$
+ \eta_1 \Big(str\int J(z)\Big[\frac{\delta \phi}{\delta J(z)},
\frac{\delta \psi}{\delta J(z)}\Big]+
e_0 str\int\partial_z\Big(\frac{\delta \phi}{\delta J(z)}\Big)
\frac{\delta \psi}{\delta J(z)}\Big)
$$
Let now $e_0=1,e_1=0$. Considering two cases: $\eta_1=0$ or $\eta_0=0$
we obtain two types of Poisson brackets:

\begin{eqnarray}\label{eq:puasson}
&&\{\phi,\psi\}_1(J)=- str\int C\Big[\frac{\delta \phi}{\delta J(z)},
\frac{\delta \psi}{\delta J(z)}\Big]
\\
&&\{\phi,\psi\}_2(J)= str\int J(z)\Big[\frac{\delta \phi}{\delta J(z)},
\frac{\delta \psi}{\delta J(z)}\Big]+
 str\int\partial_z\Big(\frac{\delta \phi}{\delta J(z)}\Big)
\frac{\delta \psi}{\delta J(z)}\nonumber
\end{eqnarray}
So, let us summarize, what was obtained.
Two Poisson brackets (\ref{eq:puasson}) are defined on 
the space $H_{C}({\cal G})$ of the following operators: 
\begin{equation}\label{eq:h_oper}
\partial_x+J(x)+\lambda C.
\end{equation}
 These Hamiltonian structures are invariant under the mentioned above
nonabelian gauge transformations of the operators
(\ref{eq:h_oper}), for which the following condition is satisfied: 
$gCg^{-1}=C$, $g\in G$.

Now let us move to the ${\rm osp}(1|2)$ case. 
One can reduce  
$H_{C}({\rm osp}(1|2))$ to the subspace 
$
H^{constr}_{C}({\rm osp}(1|2))=\{ {\cal L}=\partial_x+\alpha (x)v_-+w(x)X_-
+q(x)h+\beta(x)v_++\lambda X_- + X_+,
$
where $\alpha(x)$ is an odd element of Grassmann algebra and $q(x), w(x)$ are 
even ones\}.
Note, that we have put $C=X_-$. 
The matrix form ${\cal L}$ is:
\begin{equation}\label{eq:L-oper}
{\cal L}=\partial_x+\left(\begin{array}{ccc}
q(x) & \beta(x) & 1\\
-\alpha(x) & 0 & \beta(x)\\
\lambda+w(x) & \alpha(x) & -q(x)
\end{array}
\right)
\end{equation}
Maximal group of gauge transformations, preserving $X_-=C$, 
and therefore the form of operator ${\cal L}$ and the brackets 
(\ref{eq:puasson}) is $G_-=\{\exp(\beta(x)v_-+p(x)X_-)\}$
where $\beta(x)$ - 
odd and $ p(x)$ - even .
One can consider the factorspace $H^{constr}_{X_-}({\rm osp}(1|2))/G_-=
H^{red}({\rm osp}(1|2))$. We obtain the Hamiltonian projec\-ti\-on
$\pi:H^{constr}_{X_-} \to H^{red}_{X_-}$. It is Hamiltonian 
in the sence, that $\pi$ preserves brackets (\ref{eq:puasson}).\\
Hamiltonian reduction of such a type is called the Drinfeld-Sokolov reduction
for the superalgebra ${\rm osp}(1|2)$ [12].
Every point of the manifold
$H^{red}_{X_-}({\rm osp}(1|2))$ corresponds to the differential operator
\begin{equation}
{\cal L}=\partial_x+\left(\begin{array}{ccc}
0 & 0 & 1\\
-\varepsilon(x) & 0 & 0\\
u(x)+\lambda & \varepsilon(x) & 0
\end{array}
\right)
\end{equation}
where $u(x)$, $\varepsilon(x)$ are even and odd elements of Grassmann
algebra correspon\-ding\-ly.
Really, we have taken one element from each orbit of the
group $G_-$, acting on $H^{constr}_{X_-}({\rm osp}(1|2))$.
Now, let's consider the linear problem ${\cal L}\Psi=0$, where
$$
\Psi=\left(\begin{array}{c}
\psi_1(x)\\
\psi_2(x)\\
\psi_3(x)
\end{array}
\right)
$$
with $\psi_1(x), \psi_3(x)$ lying in the
even part of Grassmann algebra and $\psi_2(x)$ lying in the odd part.
This problem reduces to the scalar one:
\begin{equation}\label{eq:scal}
L\psi_1(x):=
(-\partial^2_x+\lambda+u(x)-\varepsilon(x)\partial^{-1}\varepsilon(x))
)\psi_1(x,\lambda)=0
\end{equation}
If we consider the problem 
${\cal L}'\Psi=0$, где $g{\cal L}g^{-1}={\cal L}'$, we obtain 
(\ref{eq:scal}) again, because $\psi_1$ is invariant 
under the transformations of $G_-$ element.
Really,
\begin{eqnarray}
&&g^{-1}\Psi=\left(\begin{array}{ccc}
1 & & \\
-\beta(x)& 1 & \\
p(x) & \beta(x) & 1
\end{array}
\right)
\left(\begin{array}{c}
\psi_1(x)\\
\psi_2(x)\\
\psi_3(x)
\end{array}
\right)=
\left(\begin{array}{c}
\tilde{\psi}_1(x)\\
\tilde{\psi}_2(x)\\
\tilde{\psi}_3(x)
\end{array}
\right)=
\tilde{\Psi}
\\
&& \Rightarrow g{\cal L}g^{-1}\Psi=0 \Leftrightarrow {\cal L}\tilde{\Psi}=0
\Leftrightarrow (\ref{eq:scal})\nonumber
\end{eqnarray}
It seems that $H^{red}_{X_-}({\rm osp}(1|2))$ 
can be idetified with operators 
$ L=-\partial^2_x+\lambda+u(x)-\varepsilon(x)\partial^{-1}\varepsilon(x)$.
But we can't do this because we can loose the part of
the solutions of the matrix linear problem ${\cal L}\Psi=0$.
For example, in the case of 1-dimensional Grassmann algebra operator $L$ 
coincides with the Sturm-Liouville operator and does not take into account 
the solution of the following form:
\begin{equation}
\Psi=
\left(\begin{array}{c}
0\\
\theta c\\
0
\end{array}
\right),
\end{equation}
where $ c$ is a constant and $\theta$ is an odd basis element of Grassmann
algebra.
Let's now rewrite the Poisson brackets (\ref{eq:puasson}) on the manifold 
$H^{red}_{X_-}({\rm osp}(1/2))$ in terms of its parameters 
$u(x), \varepsilon(x)$ :
\begin{eqnarray}\label{brac1}
\{u(x),u(y)\}_2&=&
\frac{1}{2}\Big(\delta'''(x-y)-2u'(x)\delta(x-y)-4u(x)\delta'(x-y)\Big)
\nonumber\\
\{u(x),\varepsilon(y)\}_2&=&
-\frac{1}{2}\Big(3\varepsilon(x)\delta'(x-y)+\varepsilon'(x)\delta(x-y)\Big)
\\
\{\varepsilon(x),\varepsilon(y)\}_2&=&
-\frac{1}{2}\Big(\delta''(x-y)-u(x)\delta(x-y)\Big)\nonumber
\end{eqnarray}
\begin{eqnarray}\label{brac2}
\{u(x),u(y)\}_1&=&-2\delta'(x-y)\nonumber\\
\{u(x),\varepsilon(y)\}_1&=&0\\
\{\varepsilon(x),\varepsilon(y)\}_1&=&\frac{1}{2}\delta(x-y)\nonumber
\end{eqnarray}
Thus, $u$, $\varepsilon$ generate the superconformal algebra with
the brackets (\ref{brac1}). Poisson brackets (\ref{brac2})
are the linearization of (\ref{brac1})
.
\subsection{Infinite family of Hamiltonians.\\ 
The super-KdV equation.}

We can rewrite (13) in such a way:
\begin {equation}
 (-\partial^2_x+u(x)-\varepsilon(x)\partial^{-1}\varepsilon(x))
\psi(x,k)=k^2 \psi(x,k)\quad
(k^2=-\lambda).
\end  {equation}
Using the following ansatz: 
$$\psi(x,k)=
exp(-ikx+\sum^\infty_{n=0}f_n(x)(2ik)^{-n})
$$
substituting it in (18), we obtain the set of equations on $f_n(x)$.
Changing $\partial^{-1}$ by $\int^x_{0}$ and
$f_n(x)$ by $\int^x_{0}\sigma_n(y) dy$, we obtain that
$\int^{2\pi}_{0}\sigma_n(y)dy=H_n({\cal L})$ -
form an involutive family of gauge invariant functionals of ${\cal L}$.
Here are the first three of them:
\begin{eqnarray*}
H_1&=&\int u(x)dx\\
H_2&=&\int(u^2(x)-4\varepsilon(x)\varepsilon'(x))dx\\
H_3&=&\int((u')^2+2u^3-16\varepsilon'(x)\varepsilon''(x)-
24\varepsilon(x)\varepsilon'(x)u(x))dx
\end{eqnarray*}
From a Drinfeld-Sokolov theory we obtain the following property:
$$
\{H_{i+1},\phi\}_1=\{H_i,\phi\}_2.
$$
Now, let's write down equations of motion for $u(x,t)$, $\varepsilon(x,t)$
with the Ha\-mil\-toni\-an $H_2$ and the bracket $\{,\}_2$: 
\begin{eqnarray}
u_t&=&-u_{xxx}+6uu_x-12\varepsilon\varepsilon_{xx},\\
\varepsilon_t&=&-4\varepsilon_{xxx}+6u\varepsilon_x +3u_x\varepsilon.
\nonumber
\end{eqnarray}
System of equations (19) is called a super-KdV equation.
At first it was obtained in [4], [5].
Moreover, super-KdV system can be obtained from the 
$\cal{L}$, $\cal{M}$ pair, that is, as a compatibility conditions 
for the two equations:
$$
{\cal L}\psi=0
$$
$$
\partial_t\psi=\cal{M}\psi
$$
where
\begin{equation}
\cal{M}= \left(\begin{array}{ccc}
-u_x&-4\varepsilon_x&4\lambda-2u\\
2\varepsilon u-4\varepsilon\lambda-4\varepsilon_{xx}&0&-4\varepsilon_x\\
4\epsilon\varepsilon_x+u_{xx}-2u^2+2u\lambda+4\lambda^2&-2\varepsilon u+
4\varepsilon\lambda+4\varepsilon_{xx}&u_x\\
\end{array}\right).
\end{equation}

\section{Inverse Scattering Method for the super-KdV
equation}
In this part of the article the super-KdV equation 
is considered from a point of view of the
inverse scattering method. At first the direct scattering problem  
is analyzed and the relations between the elements of transfer matrix 
$T(k)$ are found.
Then, in the case of 1-dimensional Grassmann algebra we consider
the inverse problem. 
At the end the explicit solitonic solutions are introduced.

\subsection{Direct Problem}
The super-KdV equation can be obtained as a compatibility condition for 
two equations:
$$
{\cal L}\psi=0
$$
$$
\partial_t\psi=\cal{M}\psi
$$
In this subsection we analyze the linear problem ${\cal L}\psi=0$.
Let's transform it. First of all we consider the group element $U=exp(ikX_-)$
and using it we make a similarity transformation $U{\cal L}U^{-1}$,
thus obtaining the operator:
\begin{equation} \label{L'}
{\cal L'}=\partial_x+
\left(\begin{array}{ccc}
-ik&0&1\\
-\varepsilon&0&0\\
u&\varepsilon&ik\\
\end{array}\right)
\end{equation}
Then, using the matrix N:
\begin{equation}
N=
\left(\begin{array}{ccc}
1&0&-(2ik)^{-1}\\
0&1&0\\
0&0&(2ik)^{-1}\\
\end{array}\right)
\end{equation}
one can bring the constant part of ${\cal L'}$ to the diagonal form:
\begin{eqnarray*}
\tilde{{\cal L}}(x)=
\left(\begin{array}{ccc}
-u(2ik)^{-1}&-\varepsilon(2ik)^{-1}&-u(2ik)^{-1}\\
-\varepsilon&0&-\varepsilon\\
u(2ik)^{-1}&\varepsilon(2ik)^{-1}&u(2ik)^{-1}\\
\end{array}\right)+
\left(\begin{array}{ccc}
-ik& & \\
& 0&\\
 & & ik
\end{array}\right)+
\partial_x 
\end{eqnarray*}
We can rewrite the corresponding linear problem as follows:
$$\partial_x \Psi=ikh\Psi + Q(x,k)\Psi,$$ where
\begin{equation} \label{Q}
Q(x,k)=
\left(\begin{array}{ccc}
u(2ik)^{-1}&\varepsilon(2ik)^{-1}&u(2ik)^{-1}\\
\varepsilon&0&\varepsilon\\
-u(2ik)^{-1}&-\varepsilon(2ik)^{-1}&-u(2ik)^{-1}\\
\end{array}\right)
\end{equation}
Let's consider the following matrix solutions (Jost solutions):

\begin{eqnarray} \label{Phi}
&&\Phi^+(x,k)=
\left(\begin{array}{ccc}
e^{ikx}&0&0\\
0&1&0\\
0&0&e^{-ikx}\\
\end{array}\right)\qquad x\to  \infty,\\
&&\nonumber\\
&&\Phi^-(x,k)=
\left(\begin{array}{ccc}
e^{ikx}&0&0\\
0&1&0\\
0&0&e^{-ikx}\\
\end{array}\right)\qquad x\to - \infty\nonumber
\end{eqnarray}
This matrix valued functions are constructed in such a way 
that $N^{-1}\Phi^{\pm}N$ 
are the elements of the group ${\rm Osp}(1|2)$.
Let $\Phi^+(x,k)T(k)=\Phi^-(x,k)$, where
\begin{equation}
T(k)=
\left(\begin{array}{ccc}
a(k)&\gamma(k)&b(k)\\
\xi(k)&f(k)&\delta(k)\\
c(k)&\eta(k)&d(k)\\
\end{array}\right)
\end{equation}
is a transfer matrix. 
One can find relations between elements of $T(k)$. 
Really, defining the matrix $P$:
\begin{equation}
P=
\left(\begin{array}{ccc}
0&0&1\\
0&1&0\\
1&0&0
\end{array}\right)
\end{equation}
it is easy to see, that if $\Psi(x,k)$ is a solution of (\ref{Q}),
then $P\Psi(x,k)P=\tilde{\Psi}(x,k)$ 
satisfies the following equation:
$$
\partial_xP\Psi P= -ikhP\Psi P+Q^* P\Psi P
$$
where * means the reversing of the sign of $k$ (if $k\in {\bf R}$, then it
is the complex conjugation):
$$
\tilde{\Psi}(x,k)=\Psi(x,-k)=\Psi^*(x,k).
$$
Thus $T^*=PTP$ and we can reduce
the transfer matrix [7],[8]: 
\begin{equation}
T(k)=
\left(\begin{array}{ccc}
a&\gamma&b\\
\xi&f&\delta\\
c&\eta&d\\
\end{array}\right)=
\left(\begin{array}{ccc}
a&\gamma&b\\
\bar{\delta}&f&\delta\\
\bar{b}&\bar{\gamma}&\bar{a}\\
\end{array}\right)
\end{equation}
that is $\xi=\bar{\delta},\ d=\bar{a},\ \eta=\bar{\gamma}
,\ f=\bar{f}$ (bar means the same as *).
Moreover, we know that $N^{-1}TN\in {\rm Osp}(1|2)$.
In such a way we have found the constraints:
\begin{eqnarray*}
&&f=1+2ik\bar{\gamma}\gamma\\
&&f(a\bar{a}-b\bar{b})=1\\
&&\delta=2ik(\bar{\gamma}b-\gamma\bar{a})
\end{eqnarray*}
In the case of 1-dimensional Grassmann algeba
they have the following form:
\begin{eqnarray*}
f=1,\quad  a\bar{a}-b\bar{b}=1,\quad 
\delta=2ik(\bar{\gamma}b-\gamma\bar{a})
\end{eqnarray*}
After this we consider the factorization of the matrix $T(k)$: $T^+=TT^-$,
where
\begin{equation}
T^+(k)=
\left(\begin{array}{ccc}
1 & -\delta(2ik)^{-1} & b\\
0 &  \bar a  & \delta\\
0 & 0 & \bar a\\
\end{array}\right),\qquad
T^-(k)=
\left(\begin{array}{ccc}
\bar a & 0 & 0\\
-(2ik)\bar \gamma &\bar a  & 0\\
-\bar b & -\bar \gamma & 1\\
\end{array}\right)
\end{equation}
One can construct the matrix-valued function:
\begin{equation}
\Psi^+(x,k) =\Phi^+(x,k)T^+(k)e^{-ikxh}=\Phi^-(x,k) T^-(k)e^{-ikxh} 
\end{equation}
Asymptotics of 
it's diagonal components have the folowing
expression:
\begin{equation}
\pi_0 T^+=
\left(\begin{array}{ccc}
1 & 0 & 0\\
0 & \bar a & 0\\
0 & 0 & \bar a\\
\end{array}\right)
\end{equation}
$$x \to \infty$$ 
\begin{equation}
\pi_0 T^-=
\left(\begin{array}{ccc}
\bar a & 0 & 0\\
0 & \bar a & 0\\
0 & 0 & 1\\
\end{array}\right)
\end{equation}
$$x \to -\infty$$
$\Psi^+(x,k)$ satisfies the integral equation:
\begin{eqnarray}
\Psi^+(x,k)=\pi_0 T^+ &+&\int^x_{-\infty}e^{ikad_h(x-y)} \pi_+(Q(y,k)\Psi^+(y,k))dy\\
&-&\int^{\infty}_x e^{ikad_h(x-y)} (\pi_0+\pi_-)(Q(y,k)\Psi^+(y,k))dy\nonumber
\end{eqnarray}
where $\pi_0$ is the projection on the diagonal part of the corresponding 
matrix,
and $\pi_\pm$ are the projections on
the strict upper triangular and strict lower triangular
parts correspondingly.
Let's write the integral equation for the first
column of $\Psi^+(x,k)$:
\begin{eqnarray*}
&&\Psi^+(x,k)=
\left(\begin{array}{c}
\Psi_{11}\\
\Psi_{21}\\
\Psi_{31}\\
\end{array}\right)
=\left(\begin{array}{c}
1\\
0\\
0\\
\end{array}\right)+\\
&&\\
&&
\left(\begin{array}{l}
-(2ik)^{-1}\int_x^{\infty}((\Psi_{11}(y,k)+\Psi_{31}(y,k))u(y)-
\varepsilon(y) \Psi_{21}(y,k)) dy\\
\\
-\int_x^{\infty}e^{-ik(x-y)}(\varepsilon(y)(\Psi_{11}(y,k)+\Psi_{31}(y,k)))dy\\
\\
(2ik)^{-1}\int_x^{\infty}e^{-2ik(x-y)}((\Psi_{11}(y,k)+
\Psi_{31}(y,k))u(y)-\varepsilon(y) \Psi_{21}(y,k)) dy\\
\end{array}\right).
\end{eqnarray*}
We derive from this the relation 
for the first column of the matrix
Jost solution $\Phi^{+(1)}(x,k)=\Psi^{+(1)}(x,k)e^{ikx}$ and, correspondingly,
$$N^{-1}\Psi^{+(1)}(x,k)=\tilde m_+(x,k),
$$ 
where
\begin{equation} \label{tildem}
\tilde m_+(x,k)=\left(\begin{array}{c}
1+\int_x^{\infty} dy \frac {e^{-2ik(x-y)}-1} {2ik} 
(\tilde m^1_+(y,k)u(y)-\epsilon(y)\tilde m^2_+(y,k))\\
\\{}
\ -\int_x^{\infty} dy e^{-ik(x-y)}\epsilon(y)\tilde m_+^1(y,k) \\
\\{}
  \int_x^{\infty} dy e^{-2ik(x-y)} (u(y)\tilde m_+^1(y,k)-
\epsilon(y)\tilde m^2_+(y,k)) 
\end{array}\right).
\end{equation} 
Now we can see, that $\tilde m_+(x,k)$
allows analytical continuation in the upper half-plane of $k$ .
Analogously, considering factorization $R^-=TR^+$ and corresponding integral 
equations, one can prove that $\Phi^{-(1)}(x,k)e^{-ikx}$ is
analytic in the lower half-plane.
From $T^+, T^-$ factorization we obtain that
\begin{eqnarray} \label{Phi+}
\Phi^{+(1)}(x,k)e^{-ikx}&=&\bar a(k) \Phi^{-(1)}(x,k)e^{-ikx} - 
2ik \bar\gamma(k)\Phi^{-(2)}(x,k)e^{-ikx}\\
&-&
\bar b \Phi^{-(3)}(x,k)e^{-ikx}\nonumber
\end{eqnarray}
Using the above reasoning, one can rewrite it in the following way:
\begin{eqnarray}
m_+(x,k) - m_-(x,k) &=& -2ik\rho(k)N^{-1}(k)\Phi^{-(2)}(x,k)e^{-ikx} \\
&-& r(k)e^{-2ikx} P(k)m_- (x,-k)\nonumber
\end{eqnarray}
where
\begin{eqnarray*}
&& m_+(x,k)=\frac {\tilde m_+(x,k)} {\bar a(k)}, \quad 
m_-(x,k)=\tilde m_-(x,k) ,
\quad \rho(k)=\frac {\bar\gamma(k)} {\bar a(k)},\\ 
&& r(k)=\frac{\bar b(k)}{\bar a(k)},\quad P(k)=N^{-1}(k)PN(-k)
\end{eqnarray*}
or :
\begin{equation}\label{m+}
m_+(x,k) - m_-(x,k) = V(k,x) m_- (x,-k) + f(x,k)
\end{equation}
where
$$
V(k,x)=- r(k)e^{-2ikx} P(k), \qquad f(x,k)= -2ik\rho(k)
N^{-1}(k)\Phi^{-2}(x,k)e^{-ikx}
$$
Moreover, the equation (\ref{tildem}) gives:
\begin{equation}\label{restore}
u(x)=2ik\partial_x\tilde m^1_+(x,k), \qquad 
\varepsilon(x)=ik\tilde m^2_+(x,k),\qquad |k|\to\infty 
\end{equation}
Thus, we can restore $u(x)$, $\varepsilon(x)$ by means of $r(k)$, 
$\rho(k)$, solving (\ref{m+}). 
However, in the general case we can say nothing about
the behaviour of
$\Phi^{-(2)}(x,k)$, so, from now on,  
we study the case of 1-dimensional Grassmann algebra, for which 
$$
\rho(k)\Phi^{-(2)}(x,k)=
\left(\begin{array}{c}
0\\
\rho(k)\\
0\\
\end{array}\right)
$$
Also, in 1-dimensional case all even elements satisfy the usual KdV
properties.
Variable $k$ may be situated whether on the real line or in 
the discrete set of points on the imaginary axis.
This happens because of the fact, that in 1-dimensional case
our $\cal L$ problem reduces to the Sturm-Liouville problem,
for which, in the case of fast decreasing potential
we have the continuous spectrum: positive real axis and 
the discrete spectrum:
the set of points on the negative real axis (see [7],[8]).
It is known that $a(k)$ has the following asymptotics:
$$
a(k)=1+O\Big(\frac{1}{k}\Big) \quad {\rm при}\quad |k|\to\infty,
$$ 
$a(k)$ is analytic in the lower half-plane and
$\bar a(k)=a(-k)$ in the upper one. Points of the discrete spectrum 
corresponds to zeros of $a(k)$ (correspondingly $\bar a(k)$) 
so, for any point of discrete spectrum $i\kappa_j$ 
there exist simple zero at this point of 
$a(k)$ ($\bar a(k)$) and if $i\kappa_j$ is situated in 
the upper half-plane, this point correspons to the simple pole
of $m_+(x,k)$.

So, if we know the so-called ``scattering data'': 
$$\{i\kappa_\ell\}, 
\{b_\ell\}, \{\gamma_\ell\}, \rho(k), r(k)$$
we can restore $u(x),\varepsilon(x)$, solving the Riemann problem
(\ref{m+}).

\subsection{Inverse problem}
Let's summarise the results obtained from study of the direct problem:
\begin{enumerate}
\item The scattering data:
$$\{i\kappa_\ell\}, 
\{b_\ell\}, \{\gamma_\ell\}, \rho(k), r(k)$$
\item Riemann problem:
\begin{equation}\label{Rm+}
m_+(x,k) - m_-(x,k) = V(x,k)m_-(x,k)+f(x,k),
\end{equation}
where 
$m_+(x,k)$ is meromorphic in the upper half-plane 
with simple poles at the points $\{i\kappa_\ell\}$, $m_-(x,k)$ 
is analytic in the lower half-plane with the symptotics:
$m_\pm(x,k)\to (1,0,0)^t$ when $|k|\to \infty$
\begin{eqnarray}
V(x,k)=-r(k)e^{-2ikx}N^{-1}(k)PN(-k), \\
f(x,k)=
\left(\begin{array}{c}
0\\
-2ik\rho(k)e^{-2ikx}\\
0\\
\end{array}\right)\nonumber
\end{eqnarray}
The relation (\ref{Rm+}) is written for $k\in{\bf R}$. 
\end{enumerate}
We want to find $m_+(x,k)$, and then, using (\ref{restore}),
we obtain $u(x)$, $\varepsilon(x)$ by the formulas:
$$
u(x)=2ik\partial_xm^1_+(x,k),\quad  \varepsilon(x)=ikm^2_+(x,k),
\quad k\to\infty
$$
The solution of (\ref{Rm+}) has the form [9]:
\begin{eqnarray}
 m(x,k)=\left(\begin{array}{c}
1\\
 0\\
0 
\end{array}\right)
&+& \sum^N_{j=1}\frac{m_j(x)}{k-i\kappa_j}\\
&+&\frac{1}{2 \pi i}\int^{+\infty}_{-\infty}dz
\frac {V(x,z)m_-(-z,x)+f(x,z)}{z-k}\nonumber
\end{eqnarray} 
$m(x,k)$ is a function, which coincides with
$m_-(x,k)$ in the lower half-plane; in the upper half-plane it is 
equal to $m_+(x,k)$ and on the real axis it has a jump (38).
Moreover,
$$ 
m_\ell(x)=\frac{\Phi^{+1}(x,i\kappa_\ell)}{\bar a'(i\kappa_\ell)} e^{\kappa_\ell x}
$$ 
(see subsection 1).
Let's consider the relation (\ref{Phi+}) at the point $i\kappa_n$:
\begin{equation}
 \Phi^{+(1)}(x,i\kappa)e^{\kappa_n x}= 2\kappa_n \bar\gamma_n\Phi^{-(2)}(x,i\kappa_n)e^{\kappa_n x}-\bar b_n 
P(i\kappa_n)\Phi^{-(1)}(x,-i\kappa_n)e^{2\kappa_n x}
\end{equation}
We deduce:
\begin{eqnarray}
&&m_n(x)= 
\frac {2\kappa_n\bar\gamma_n e^{\kappa_n x}}{\bar a'(i\kappa_n)}\Phi^{-2}(x,i\kappa_n)-
\frac{ \bar b_n e^{2\kappa_n x}}{\bar a'(i\kappa_n)}P(i\kappa_n)\cdot \\
&& \Big(\left(\begin{array}{c}
1\\
 0\\
0 
\end{array}\right)
+ i\sum^N_{j=1}\frac{m_j(x)}{\kappa_n + \kappa_j}+\frac{1}{2 \pi i}\int^{+\infty}_{-\infty}dz
\frac {V(x,z)m_-(x,-z)+f(x,z)}{z-i\kappa_n}\Big)\nonumber
\end{eqnarray} 

In such a way we obtain the following equations for 
the first two components of $m(x,k)$ (we denote
them 
$R(x,k)$ and $\Theta(x,k)$): 
\begin{eqnarray}\label{RTheta}
R_n(x)&=&-\frac{ \bar b_n e^{2\kappa_n x}}{\bar a'(i\kappa_n)}\Big(1+ 
i\sum^N_{j=1}\frac{R_j(x)}{\kappa_n + \kappa_j} -\frac{1}{2 \pi i}
\int^{+\infty}_{-\infty}dz\frac {r(z)R(x,-z)e^{-2izx}}{z-i\kappa_n}\Big) 
\nonumber\\
R(x)&=&1+\sum^N_{j=1}\frac{R_j(x)}{k - i\kappa_j}-
\int^{+\infty}_{-\infty}dz\frac {r(z)R(x,-z)e^{-2izx}}{z-(k+i0)}
\nonumber\\
\Theta_n (x)&=&\frac {2\kappa_n\bar\gamma_n e^{\kappa_n x}}{\bar a'(i\kappa_n)}-
-\frac{ \bar b_n e^{2\kappa_n x}}{\bar a'(i\kappa_n)}
\Big(i\sum^N_{j=1}\frac{\Theta_j(x)}{\kappa_n + \kappa_j} 
\\
&-&\frac{1}{2 \pi i}
\int^{+\infty}_{-\infty}dz\frac {r(z)\Theta(x,-z)e^{-2izx}+2iz\rho(z)e^{-izx}}{z-i\kappa_n}\Big)
\nonumber\\
\Theta(x)&=&\sum^N_{j=1}\frac{\Theta_j(x)}{k - i\kappa_j}-\frac{1}{2 \pi i}
\int^{+\infty}_{-\infty}dz\frac {r(z)\Theta(x,-z)e^{-2izx}+2iz\rho(z)e^{-izx}}{z-(k+i0)}\nonumber
\end{eqnarray}
Using these formulas we obtain the expressions for the potentials:

\begin{eqnarray}\label{potential}
&&u(x)=\partial_x \Big(2i\sum^N_{j=1} R_j(x)+\frac{1}{\pi}
\int^{+\infty}_{-\infty}dz(r(z)R(x,-z)e^{-2izx})\Big)
\\
&&\varepsilon(x)=i\sum^N_{j=1}\Theta_j(x)+\frac{1}{2\pi}
\int^{+\infty}_{-\infty}dz\Big(r(z)\Theta(x,-z)e^{-2izx}+2iz\rho(z)e^{-izx}\Big)
\nonumber
\end{eqnarray}
Now, let's recall the equation $\partial_t\Psi={\cal M}\Psi$ from the
beginning of subsection 1. 
One can include the dependence on $t$ in $T$, using the
relation $\Phi^-=\Phi^+ T$ with the
$x$-asymptotics of $\Phi^-$ и, $\Phi^+$ unchanged, that is:
$$
T(k,t)=e^{4ik^3th}T(k)e^{-4ik^3th}.
$$
Then, the equations describing the dynamics of coefficients
of transfer matrix can be written as follows:
\begin{eqnarray*}
\dot{\bar a}(k,t)&=&0\\
\dot{\bar\gamma}(k,t)&=&-4ik^3{\bar\gamma}(k,t)\\
\dot{\bar b}(k,t)&=&-8ik^3{\bar b}(k,t)
\end{eqnarray*}
The expressions giving the potentials $u(x,t)$, $\varepsilon(x,t)$
coincide with (\ref{potential}), with
the included dynamics of transfer matrix coefficients. The 
equations (\ref{RTheta}) are explicitly solvable in the case
of reflectionless potentials, when $b(k)=0$ and
 $\rho(k)=0$. In this case the integral equations
reduce to the algebraic ones:
\begin{eqnarray*}
R_n(x)&=&-\frac{ \bar b_n e^{2\kappa_n x}}{\bar a'(i\kappa_n)}\Big(1+ 
i\sum^N_{j=1}\frac{R_j(x)}{\kappa_n + \kappa_j}\Big)  
\\
\Theta_n (x) &=& \frac {2\kappa_n\bar\gamma_n e^{\kappa_n x}}{\bar a'(i\kappa_n)}
-i\frac{ \bar b_n e^{2\kappa_n x}}{\bar a'(i\kappa_n)}
\sum^N_{j=1}\frac{\Theta_j(x)}{\kappa_n + \kappa_j}
\\
u(x)&=&2i\partial_x \sum^N_{j=1} R_j(x)
\\
\varepsilon(x)&=&i\sum^N_{j=1}\Theta_j(x)
\end{eqnarray*}
As an example one can calculate 1-soliton solution
(only one point of the discrete spectrum $\lambda_\kappa=\kappa^2$).
We obtain:
\begin{equation}
u(x,t)=-\frac{c}{2 ch^2\Big(\frac{\sqrt {c}}{2}(x-ct)\Big)},
\qquad
\varepsilon(x,t)=-\frac{\alpha}{ ch\Big(\frac{\sqrt {c}}{2}(x-ct)\Big)}
\end{equation}
where $c=4\kappa^2,\bar b_1(0)=-1$  and $\alpha$  is
any odd constant element of the Grassmann algebra.

Actually, using ISM, we can solve KdV by means of 
implicit changing of variables
$u(x)$, $\varepsilon(x)\to s(0)$ (the scattering data). In terms of new 
variables the equations of motion $\dot{s}=F(s)$ 
are trivially solvable diferential equations. Inverse transformation
$$s(t)\to \varepsilon(x,t),u(x,t)
$$ 
with the use of the integral equations (43),(44) 
and the evolution of
the scattering data gives the solution for super-KdV system.
Thus, here is the scheme for solving the Cauchy problem [6]-[8]:

\vspace{5mm}

\begin{tabular}{ccccccc}
 &I& &II&&III&\\
$\varepsilon(x,0),u(x,0)$&$\to$&$s(0)$&$\to$&$s(t)$&$\to$&$u(x,t),\varepsilon(x,t)$\\
\end{tabular}

\vspace{5mm}
At first we obtain from $u(x,0),\varepsilon(x,0)$ 
two functions $\rho(k),r(k)$ on the half-line (by virtue of
relations $\bar r(k)=r(-k), \bar\rho(k)=\rho(-k)$) 
and the sets $\{i\kappa_j\}, \{\bar\gamma_j\}, \{\bar b_j\}$.
Then, turning on the dynamics 
\begin{eqnarray*}
\dot{\bar\gamma}(t)&=&-4ik^3\gamma(t)\\
\bar b(t)&=&-8ik^3\bar b(t)\\
\dot{\bar a}(t)&=&0
\end{eqnarray*} 
we can restore the potentials, using the equations
(\ref{RTheta}),(\ref{potential}). 
 \section{Appendix}

\subsection{Grassmann Algebra}
\underline{Definition} 
[13],[14] The finite dimensional Grassmann algebra $\Lambda(n)$ 
of order $n$ 
is an algebra with $n$ generators: 
$1,\Theta_1,\dots,\Theta_n$, satisfying the anticommutati\-vi\-ty
property
$\Theta_i\Theta_j+\Theta_j\Theta_i=0 \ $.
The element of this algebra is called even (odd), if in the decomposition
of this element 
$$\eta=\sum_{m\geq 0}
\sum_{i_1<\dots <i_m}\eta_{i_1\dots i_m}\Theta^{i_1}\dots \Theta^{i_m}
$$
each number $m$ is even (odd). As vector space $\Lambda(n)$
splits int two subspaces
$\Lambda(n)=\Lambda(n)_{\bar 0}\oplus\Lambda(n)_{\bar 1}$, where 
 $\Lambda(n)_{\bar 0}$ is a linear space of even elements and 
$\Lambda(n)_{\bar 1}$ is a linear space of odd elements . 
\subsection{Lie superalgebra $\bf{\rm\bf osp}(1|2)$}
\underline{Definition} 
[13], [14]Lie superalgebra (or $Z_2$-graded Lie algebra) 
is a real or complex $Z_2$-graded space with fixed parity 
$\alpha:{\cal G}\to  Z_2$, such that
 ${\cal G}={\cal G}_{\bar 0}\oplus {\cal G}_{\bar 1}$, where 
$\alpha({\cal G}_{\bar 0})=0$, $\alpha({\cal G}_{\bar 1})=1$, 
where the bilinear map $[,]$ is defined, 
such that for homogeneous elements (that is elements,
lying in ${\cal G}_{\bar 0}$ or in ${\cal G}_{\bar 1}$)  
the following properties are valid:
\begin{eqnarray}
&&\alpha ([x,y])=\alpha(x)+\alpha(y),\\
&&[x,y]=(-1)^{\alpha(x)\alpha(y)+1}[y,x],\\
&&[x,[y,z]](-1)^{\alpha(x)\alpha(z)}+[z,[x,y]](-1)^{\alpha(z)\alpha(y)}
\nonumber\\
&&+
[y,[z,x]](-1)^{\alpha(y)\alpha(x)}=0.\label{term}
\end{eqnarray}
As an example we consider Lie superalgebra ${\rm osp}(1|2)$:
\begin{eqnarray*}
&&[h,X^\pm]=\pm 2X^\pm \quad [h,v^\pm]=\pm v^\pm \quad [X^+,X^-]=h\\
&&[v^+,v^-]=-h \quad [X^\pm,v^\mp]=v^\pm \quad [v^\pm,v^\pm]=\pm 2X^\pm
\end{eqnarray*}
$v^\pm$ are odd; $h,X^\pm$ are even.\\
Matrix realization:
\begin{eqnarray*}
&&h=
\left(\begin{array}{ccc}
1 & 0 & 0\\
0 & 0 & 0\\
0 & 0 & -1\\
\end{array}\right)\quad
v_-= 
\left(\begin{array}{ccc}
0 & 0 & 0\\
-1 & 0 & 0\\
0 & 1 & 0\\
\end{array}\right)\quad 
v_+=
\left(\begin{array}{ccc}
0 & 1 & 0\\
0 & 0 & 1\\
0 & 0 & 0\\
\end{array}\right)\\
&&\\
&&X^+= 
\left(\begin{array}{ccc}
0 & 0 & 1\\
0 & 0 & 0\\
0 & 0 & 0\\
\end{array}\right)\quad 
X^-=
\left(\begin{array}{ccc}
0 & 0 & 0\\
0 & 0 & 0\\
1 & 0 & 0\\
\end{array}\right) 
\end{eqnarray*}

\subsection{Supergroups. Supergroup $\bf{\rm\bf osp}(1|2)$}

To construct the supergroup by means of superalgbra, we need to
consider the exponential map, as in usual case.
The difference is the following: you need to attach
to each generator an element of Grassmann algebra
of the same parity (the index of the exponent should be even).

Example: supergroup ${\rm Osp}(1|2)$ [13], [14]

This group is generated by the matrices 
(in the defining representations) of the following form:

$$
u=
\left(\begin{array}{ccc}
a & \alpha & b\\
\beta & f & \delta\\
c & \gamma & d\\
\end{array}\right), 
$$
with $uJu^{st}=J$, where

$$
J=
\left(\begin{array}{ccc}
0 & 0 & 1\\
0 & 1 & 0\\
-1 & 0 & 0\\
\end{array}\right), 
$$
а
$$
u^{st}=
\left(\begin{array}{ccc}
a & -\beta & c\\
\alpha & f & \gamma\\
b & -\delta & d\\
\end{array}\right), 
$$
is a supertransposed matrix. Supertrace of the matrix $u$ 
is defined by the expression $f-a-d={\rm str}u$.\\
The properties of the supertrace:
\begin{eqnarray*}
&&{\rm str}(MN)=(-1)^{\alpha(M)\alpha(N)}{\rm str}(NM)\\
&&{\rm str}(M+N)={\rm str}M+{\rm str}N\\
&&{\rm str}(M^{st})={\rm str}M
\end{eqnarray*}

\subsection{The notations, used in the text}

We use the following notations:
the Greek letters denote odd elements of Grassmann algebra,
the Latin letters denote even elements.

\end{document}